\documentstyle[twocolumn,prl,aps,floats,epsfig]{revtex}

\begin{document} \draft
\twocolumn[\hsize\textwidth\columnwidth\hsize\csname@twocolumnfalse\endcsname

\title{Atomic scale imaging and spectroscopy of the V$_{2}$O$_{3}$
(0001)-surface: bulk versus surface effects}
\author{M. Preisinger, J. Will, M. Klemm, S. Klimm, and S. Horn}
\address{Lehrstuhl f\"{u}r Experimentalphysik II, Universit\"{a}t Augsburg, 86135
Augsburg, Germany} \date{\today} \maketitle

\begin{abstract} We present atomic scale images of a V$_{2}$O$_{3}$
(0001)-surface, which show that the surface is susceptible to reconstruction by
dimerization of vanadium ions. The atomic order of the surface depends
sensitively on the surface preparation. Scanning tunneling spectroscopy proves
a dimerized surface has a gap in the electronic density of states at the Fermi
energy, while a surface prepared by sputtering and successive annealing shows
no dimerization and no gap. Photoemission spectra depend sensitively on the
surface structure and are consistent with scanning tunneling spectroscopy data.
The measurements explain inconsistencies in photoemission experiments performed
on such oxides in the past.\end{abstract}


\vskip2.0pc ]

\narrowtext

Despite intense efforts the field of electronically correlated materials still
poses major challenges to a theoretical understanding \cite{Imada98}. An
excellent example for the richness of the problem is the transition metal oxide
V$_{2}$O$_{3}$, which shows a complex phase diagram with paramagnetic
metal~(PM), paramagnetic insulator~(PI) and antiferromagnetic insulator~(AFI)
regions as function of temperature and pressure \cite{McWhan70}. Recently,
detailed predictions for the spectral functions in the PM and PI phase as
measured by photoemission have been put forward \cite{Held01}. Numerous
photoemission studies on V$_{2}$O$_{3}$ can be found in the literature
\cite{Goering97a,Kim98,Shin90,Allen01}, but only one study \cite{Goering97a}
showed that the spectra were taken from an atomically ordered surface by
utilizing low energy electron diffraction (LEED). Since the various published
spectra differ with respect to their spectral weight at the Fermi energy
$\epsilon_{F}$ \cite{Allen01} a comparison with theory is ambiguous. Moreover,
surface states in photoemission spectra \cite{Maiti01} and surface
reconstruction \cite{Goering97b,Murray92} have been found in other oxides. To
shed light on the inconsistencies of photoemission data on V$_{2}$O$_{3}$ and
in view of the importance of reliable data to assess the validity of
theoretical approaches, measurements on rigorously characterized surfaces are
essential. Since V$_{2}$O$_{3}$ shows structural instabilities and is very
sensitive to oxygen stoichiometry, the perturbation introduced to the lattice
by a surface can be expected to render the surface electronic properties
different to that of the bulk. To investigate a possible surface reconstruction
in V$_{2}$O$_{3}$ and its effect on the photoemission spectra, we have
performed scanning tunneling microscopy (STM), scanning tunneling spectroscopy
(STS) and photoemission (PES) measurements on a V$_{2}$O$_{3}$ (0001)-surface
at room temperature. We show that the V$_{2}$O$_{3}$ surface can be prepared in
an atomically ordered state, but that two surface structures exist, depending
on the surface preparation process. The STM images of the first structure show
long range atomic order of dimerized vanadium ions with a domain structure of
roughly hexagonally shaped domains. The surface density of states exhibits a
gap at $\epsilon_{F}$ as evidenced from STS and is consistent with PES
measurements. A second surface structure shows no dimerization, STS
measurements show no gap and PES exhibits enhanced spectral weight at
$\epsilon_{F}$.

\begin{figure}
 \centerline{\epsfig{clip, file=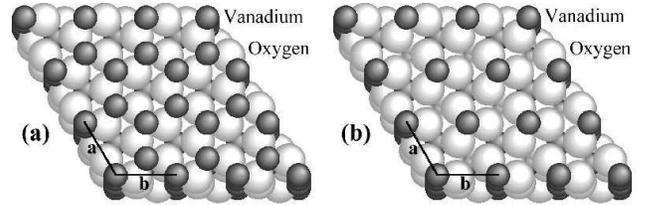, width=8.5cm}}
 \caption{(a) Honeycomb structure of vanadium terminated (0001)-surface. (b) Vanadium terminated (0001)-surface showing only vanadium ions which form a
vertical pair with a vanadium ion below. Dark spheres: Vanadium, light spheres:
Oxygen.}
 \label{Fig01}
\end{figure}

The trigonal bulk lattice structure of V$_{2}$O$_{3}$ in the metallic phase,
usually described in a hexagonal unit cell, possesses alternating vanadium and
oxygen planes in the $<$0001$>$ direction. Within the respective layers the
vanadium ions form a corrugated honeycomb lattice with a lattice constant of
0.495~nm and a vanadium-vanadium distance of 0.287~nm (Fig.~\ref{Fig01}(a)),
while the oxygen ions are arranged in a distorted hexagonal lattice with O-O
distances ranging from 0.266~nm to 0.295~nm \cite{Dernier70}. The vanadium
layer shown in Fig.~\ref{Fig01}(a) effectively exhibits two inequivalent sites,
since only half of the surface vanadium ions form a vertical pair with a
vanadium ion below. This half of vanadium ions is shown in Fig.~\ref{Fig01}(b)
constituting a simple hexagonal lattice with the same lattice constant as the
honeycomb lattice (Fig.~\ref{Fig01}(a)).

Single crystals of V$_{2}$O$_{3}$ showing specular surfaces were grown by
chemical transport. The high quality and the orientation of the crystals were
confirmed by Laue diffraction. Crystals were introduced in an Omicron UHV
system and a surface was prepared by either heating the as grown surfaces {\it
in situ} up to $750~^{\circ}$C for several minutes, or by first sputtering with
Argon in a grazing geometry followed by the same heating procedure. The
respective procedures were repeated until a LEED pattern of good quality was
observed. STM and STS measurements were performed using an Omicron variable
temperature STM (VT SPM). All STM images were recorded in the constant current
mode, using a W tip at a sample bias of $-1$~V or $+1$~V and a current of
$0.1-0.2$~nA. The topographic images show the height z(x,y) of the tip over the
sample after a plane and slope subtraction was performed. PES was carried out
using a He-discharge lamp and an Omicron AR~65 electron analyzer. Spectra are
shown after Shirley background subtraction and with the V~3d spectral weight
normalized to one.\\
\begin{figure}
 \centerline{\epsfig{clip, file=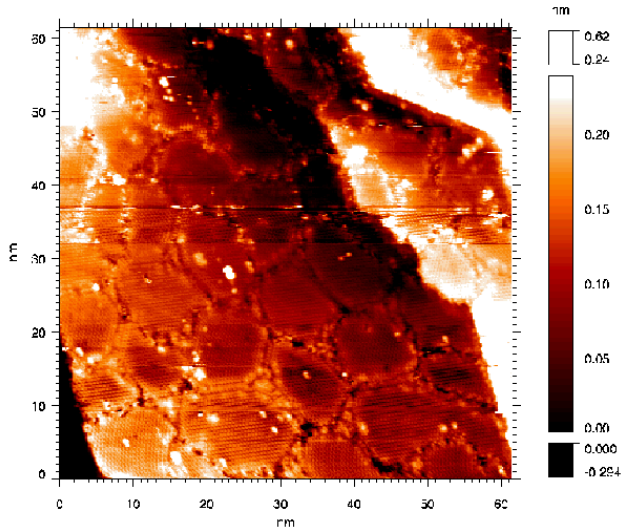, width=8.5cm}}
 \caption{Wide range STM image of V$_{2}$O$_{3}$ (0001)-surface (unsputtered),
 showing a domain like structure.
 Parameters: $I=0.1$~nA, $U_{bias}=-1$~V.}
 \label{Fig02}
\end{figure}
\begin{figure}
 \centerline{\epsfig{clip, file=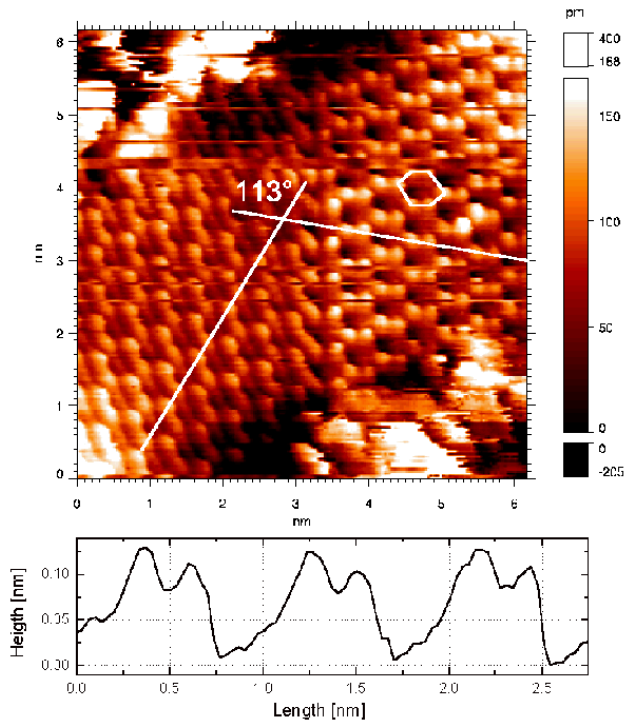, width=8.5cm}}
 \caption{Upper panel: High resolution STM image of domain boundary on unsputtered dimerized
 surface.
 Parameters: $I=0.2$~nA, $U_{bias}=+1$~V.
 Lower panel: Line scan along a row of pairs, which shows difference in vertical
 position of ions within one pair.}
 \label{Fig03}
\end{figure}
Fig.~\ref{Fig02} shows a STM topographic image over an area of $60\times60$~nm
of the V$_{2}$O$_{3}$ (0001)-surface prepared from an as grown crystal by
heating under UHV. A maximum height variation of 0.6~nm is observed over the
range displayed. The image shows a domain like structure of roughly hexagonal,
atomically ordered regions with an average domain dimension of $7-9$~nm. The
regions between adjacent domains are generally disordered and reminiscent of
domain walls or grain boundaries, with notable exceptions (see upper panel of
Fig.~\ref{Fig03}). The appearance of such a domain like structure already
indicates that the surface termination does not correspond to that expected
from the trigonal bulk structure. In the upper right corner of Fig.~\ref{Fig02}
steps of a height of $0.19-0.27$~nm are observed. This step size approximately
corresponds to the distance between two vanadium or two oxygen layers along the
$<$0001$>$ direction (0.233~nm) and suggests that the terminating layer
consists solely of either oxygen or vanadium ions. The upper panel of
Fig.~\ref{Fig03} shows two adjacent domains. Within each domain lateral pairs
of light spots are resolved, which are arranged in rows as indicated in the
figure. Here a lateral pair is identified as the neighboring spots with
shortest separation. Adjacent rows are displaced by about half the distance
between the pairs along the direction of the rows, i.e. the pairs are arranged
in a two dimensional hexagonal lattice. The periodicity is approximately
0.5~nm, which corresponds to the bulk in-plane lattice constant of the vanadium
honeycomb lattice (0.495~nm). In fact, closer inspection of the single spots,
which we identify as single atoms, show they are arranged in a distorted
honeycomb lattice, as marked in Fig.~\ref{Fig03} (upper panel). The observation
of a distorted honeycomb lattice of single atoms together with the lateral
periodicity suggests that the observed termination is a vanadium layer, in
which the V ions are dimerized. At the boundary between two domains the
orientation of the pair axis changes. The pairing axes in Fig.~\ref{Fig03}
(upper panel) form an angle of $113^{\circ}$, the angle measured for many
domains varies between $110^{\circ}$ and $130^{\circ}$. For an undistorted
honeycomb lattice (bulk vanadium layer) three equivalent directions, differing
by $120^{\circ}$, exist, along which dimerization can occur. We therefore
attribute the observed domains to a dimerization along these directions
resulting in a distortion reminiscent of that in the monoclinic AFI phase of
V$_{2}$O$_{3}$. Dimerization will most likely occur between the inequivalent
vanadium sites mentioned above, i.e. between surface vanadium ions belonging to
a vertical pair and those which do not. Further evidence for this
interpretation comes from the fact, that the two ions forming a lateral pair
differ slightly in their vertical position, as seen from the line scan in the
lower panel of Fig.~\ref{Fig03}, consistent with the respective vertical
position of the ions in the bulk structure \cite{Dernier70}.
\begin{figure}
 \centerline{\epsfig{file=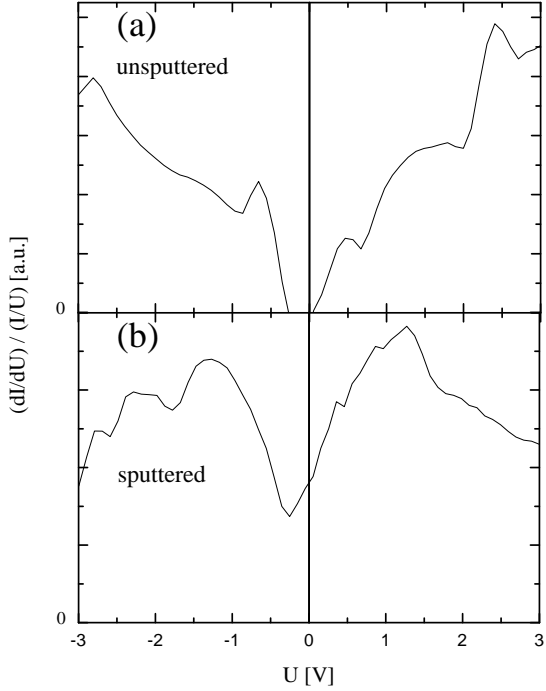, width=7.5cm}}
 \caption{(a) STS of dimerized (unsputtered) surface, exhibiting a gap in the surface DOS.
 (b) STS of mostly undimerized regions of sputtered surface, indicating metallic behavior of the surface.}
 \label{Fig04}
\end{figure}
\begin{figure}
 \centerline{\epsfig{file=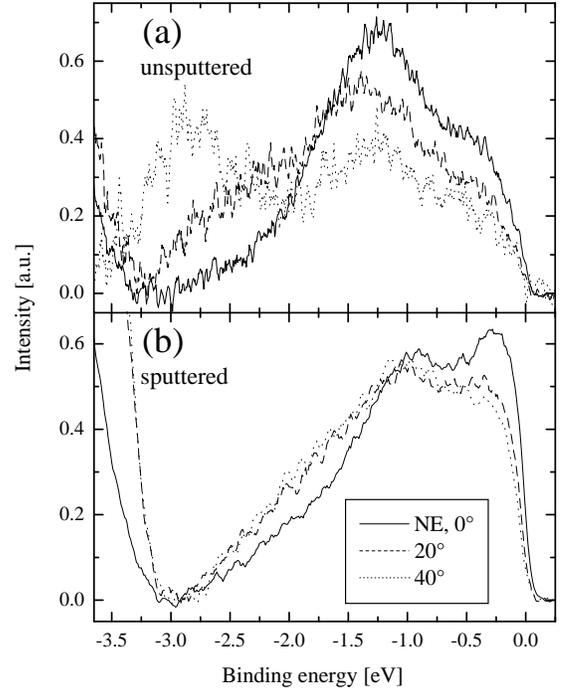, width=7.5cm}}
 \caption{PES on unsputtered dimerized (a) and sputtered (mostly undimerized) surface (b) for various
 electron take off angles.
 The dimerized surface shows considerably less spectral weight at $\epsilon_{F}$, which
 decreases with increasing take off angle.}
 \label{Fig05}
\end{figure}
A typical tunneling spectrum taken from this reconstructed surface is shown in
Fig.~\ref{Fig04}(a), in which the normalized first derivative of the tunneling
current, a measure of the surface density of states (DOS)
\cite{Feenstra87,Prietsch91}, is plotted as a function of the bias voltage. The
spectrum demonstrates a gap in the surface DOS of approximately 0.5~eV, and
displays two structures around $-0.7$~V and $-2.5$~V, suggesting maxima in the
occupied DOS at the corresponding energies. These results are consistent with
photoemission spectra, shown in Fig.~\ref{Fig05}(a), taken from the same
surface with an incident photon energy of 21.2~eV for various electron take off
angles. The spectral weight at the Fermi energy observed at normal emission
decreases with increasing take off angle, i.e. increasing surface sensitivity.
At the same time an additional peak emerges at a binding energy $E_{B}$ of
approximately $-2.7$~eV. Taking the STS results into account the spectral
weight observed in normal emission is attributed to electrons emitted from
below the surface. This contribution decreases with increasing take off angle.
The spectral weight at $E_{B}=-2.7$~eV and the corresponding fingerprint in the
STS spectrum probably originate from a state characteristic of the dimerized
surface.\\
\begin{figure}
 \centerline{\epsfig{clip, file=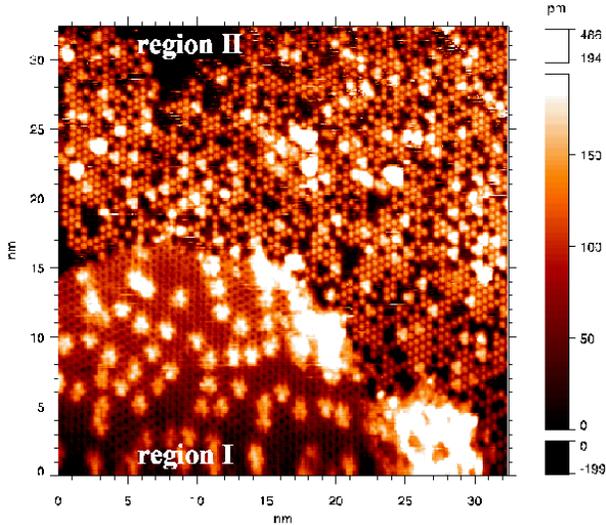, width=8.5cm}}
 \caption{STM image of boundary between dimerized (region~I) and
undimerized (region~II) regions of a sputtered surface.
 Parameters: $I=0.2$~nA, $U_{bias}=-1$~V.}
 \label{Fig06}
\end{figure}
Argon sputtering followed by successive heating produces a different surface
structure as seen from the STM image shown in Fig.~\ref{Fig06}. The figure
shows two distinguishable regions: in the lower left part (region~I) a
hexagonal lattice of dark spots is observed, interrupted by light cloud like
structures, which can in some cases be resolved into single spots forming
hexagons. Closer inspection shows the dark spots correspond to the centers of
distorted hexagons and that the structure is identical to the dimerized
structure of the unsputtered surface within a single domain. No domain
structure is found here, probably owing to the fact, that the dimerized phase
is the minority phase of this surface. Region~II, which represents the majority
phase (at least 90\% of the surface), displays a defect rich simple hexagonal
lattice of light spots. Generally the two regions are separated by a step with
the step height ranging between 0.13~nm and 0.21~nm. In some cases, as in
Fig.~\ref{Fig06}, the two regions merge without a discernable step. This
suggests the two regions both represent V ions. The hexagonal structure of
region~II can be accounted for assuming only every second V site is imaged,
compared to region~I. The periodicity of the lattice observed in region~II
(0.5~nm) is the same as that in region~I. In a straightforward interpretation
we identify the bright spots of region~II as single V ions, occupying only one
of the two inequivalent sites mentioned above, most likely sites corresponding
to a vertical vanadium pair (see Fig.~\ref{Fig01}(b)). A typical STS spectrum
of region~II (Fig.~\ref{Fig04}(b)) shows that, in contrast to region~I, no gap
exists and the structure observed at $-2.7$~V on the dimerized surface has
vanished. Correspondingly, PES (Fig.~\ref{Fig05}(b)) shows an increased
spectral weight at $\epsilon_{F}$ compared to Fig.~\ref{Fig05}(a). Still, with
increasing take off angle the spectral weight at $\epsilon_{F}$ decreases
somewhat and a structure at $E_{B}=-2.7$~eV, although much weaker than seen in
Fig.~\ref{Fig05}(a), develops. The variation of spectral weight with take off
angle probably reflects the fact that regions of dimerized vanadium ions still
exist on the sputtered surface, although the majority (undimerized) surface
phase does not exhibit a gap in the DOS.

The results presented here show the sensitivity of the V$_{2}$O$_{3}$ surface
to reconstruction. The reconstructed surface displays a gap in the DOS
resulting in a reduction of spectral weight in PES at $\epsilon_{F}$. This
effect obscures the parameter set for a theoretical description of such spectra
and shows the necessity for a thorough surface characterization to obtain a
reliable comparison between experimental and calculated spectral function. In
the case presented here, the spectra of the unreconstructed surface
(Fig.~\ref{Fig05}(b)) correspond approximately to those taken with high photon
incident energy ($>500$~eV) \cite{Allen01} (we note that using an incident
photon energy of 2000~eV further increases the spectral weight at
$\epsilon_{F}$ \cite{Woicik01} compared to Ref.~\cite{Allen01}), while spectra
taken from the reconstructed surface reflect measurements with incident
energies $<100$~eV. We attribute the slight reduction of spectral weight at
$\epsilon_{F}$ in the spectrum taken from the unreconstructed surface compared
to high incident energy measurements to the presence of a small amount of
(reconstructed) minority phase. Although V$_{2}$O$_{3}$, and, more generally,
the binary vanadium oxides, might be especially prone to surface reconstruction
in view of structural transitions taking place as a function of temperature,
the need for improved surface preparation and characterization for all the
early transition metal oxides is clear. On the other hand, the results give
hope that a carefully prepared surface can reflect the bulk properties and will
allow to study the dispersion of the quasiparticle band in the near future.

\acknowledgments This work was supported by the Deutsche Forschungsgemeinschaft
under Contract Nos. HO~522 and SFB~484.

\end{document}